\documentstyle[12pt]{article}
\begin{document}
\vspace{6\baselineskip}

\centerline{\bf{A comparison of one and two dimensional models of transonic
accretion discs }}
\centerline{\bf{around collapsed objects}}
\vskip 1.truecm
\centerline{J. PAPALOIZOU}
\vskip 0.5truecm
\centerline{and}
\vskip 0.5truecm
\centerline{E. SZUSZKIEWICZ}
\vskip 1.truecm
\centerline{\it{
Astronomy Unit, School of Mathematical Sciences }}
\centerline{\it {Queen Mary and Westfield College, University of London}}
\vskip 2.truecm
\noindent

ABSTRACT:
We construct models of the inner part of a transonic adiabatic accretion
disc assuming constant specific angular momentum taking the vertical
structure fully into account.

For comparison purposes, we construct the corresponding one dimensional
viscous disc models derived under  vertical averaging assumptions.
The conditions under which a unique location for the critical/sonic
point is obtained, given an appropriate  set of exterior boundary
conditions for these models, is also discussed.  This is not unique if the
standard '$\alpha $' prescription with viscous stress proportional
to the angular velocity gradient is used.

We use a simple model to discuss the possible limitations on the form
of the viscous stress arising from the requirement that viscous
information must travel at a finite speed. Contrary to results in the
existing literature, the viscous stress tends to be {\it increased}
rather than reduced for the type of flows we consider in which the
angular momentum and angular velocity gradients have opposite signs.
However, finite propagation effects may result
in a unique location for the sonic point.

We found good agreement between the radial flow and specific angular
momentum profiles in the inner regions of the one dimensional models
and those in the equatorial plane for corresponding two dimensional
models which may be matched for a range of $\alpha $ between 0.1
and 10$^{-4}$.

\section{Introduction}
It has long been realized that the inner regions of accretion disc
flows surrounding relatavistic objects have different
properties from those pertaining to the case when the central
object can be modelled as a Newtonian point mass.

In order to discuss such cases the approximation is often made that
the flow may be treated as Newtonian but with the modification
that the gravitational potential may be written in standard
notation following Paczy\'{n}ski and Wiita (1980) as
$$ \Psi = - {GM \over \sqrt{r^2+z^2}-r_G } , $$
where $r_G$ is the gravitational radius.
In this potential the angular momentum of circular orbits
has a maximum at $3r_G$ and so from a particle point of view
orbiting material would be expected to become unstable
and flow towards the centre attaining sonic velocities fairly
rapidly. Hence a transonic flow is expected in the central
regions ( Abramowicz and Zurek, 1981).

 Models of thick  accretion discs with steady transonic flows
were constructed by Paczy\'{n}ski (1980) who assumed that matter flowed
over the surface of an essentially hydrostatic disc, much as a star
 in a close binary system overflows its Roche lobe. Later work
assuming the accretion occurred entirely along the equatorial
plane was carried out by Paczy\'{n}ski and Abramowicz (1982) and
 R\'{o}\.{z}yczka and Muchotrzeb (1982). In this some assumptions are usually
made so that the problem is effectively reduced to a one dimensionlal one.

More recently interest has focussed on so called 'slim discs'
 ( Abramowicz, Czerny,
Lasota and Szuszkiewicz, 1988).
 These studies  reduce the
problem to a one dimensional one,  through a form of vertical
averaging even though the discs
are often geometrically thick. Radial advection is included
 and a form of the Shakura and Sunyaev (1973) prescription for the
viscosity is often  used.
 Modelling of accretion discs on this basis has been carried out  more recently
by Kato, Honma and Matsumoto (1988) and
  Chen and Taam (1993).
It is one of the purposes of this paper to construct two dimensional
 axisymmetric models of the inner region which take the vertical structure
fully
into account in order to compare the properties with those of the
one dimensional models. We make the simplifying assumptions that
the innermost regions have constant specific angular momentum and
entropy.  Apart from reasons of simplicity, the justification
 for this is that
 when the flow velocity
approaches the sound speed, the inflow or advection timescale
becomes significantly shorter than the viscous or thermal timescales
for the region of interest (see Paczy\'{n}ski and Abramowicz, 1982)
 so justifying the assumptions.
We remark that the inner regions of the one dimensional models
with viscosity included are
found to have a very slowly varying specific angular momentum
profile.

 Another issue raised recently by Narayan (1992)
is the fact that one must contrain ones treatment of viscosity so
that viscous information does not travel with supersonic velocities
(see also Chen and Taam, 1993). Narayan (1992) suggests
that the viscous stress should vanish when the flow
velocity attains the sound speed. This may have important
implications for models of transonic accretion discs esspecially
when sonic velocities are attained outside $3r_G.$

In this paper we develop a simple model
of the viscous process which has the propagation limitation built
in. We find that the viscous stress does not vanish at a sonic point
but  actually tends to be increased if the angular velocity and angular
momentum gradients are of opposite sign as occurs in the flows
discussed here. However, limitations of finite propagation speed
may affect issues associated with the uniqueness of the flow
subject to specification of appropriate boundary conditions.

 In section 2 we consider steady state accretion tori with
constant specific angular momentum  and entropy. We derive an
equation for the velocity potential in the special case of zero
vorticity. In section 3 we discuss the numerical solution of
this equation.  For the purpose of comparison of these models with one
dimensional models, we consider such models, which include
viscosity treated according to the usual Shakura and Sunyaev (1973)
 $\alpha$ prescription, derived from vertically averaged equations
 in section 4. We argue that specification of the mass
accretion rate and constant entropy of the flow coupled with
a specification that the flow have Keplerian rotation at some outer
boundary radius does not yield a unique location for the critical
point.

Further, we discuss, using a simple model, the possible
effects of the limitation that viscous information should not be
transmitted at a speed exceeding the sound velocity. We find that
if this speed coincides with the sound speed, the lack of uniqueness
in determination of the critical point discussed above may be
removed.

 In section 5 we describe our results for one and two
dimensional steady state accretion flows. When the intersection
point of the sonic surface with the equatorial plane in a two
dimensional model coincides with the critical point in a one
dimensional model and the sound velocities match there
good agreement is found for the flow parameters.

 Finally in section 6 we discuss our results.

\section {Tori with constant specific angular momentum}

In axisymmetric tori with constant specific angular momentum
 the velocity ${\bf u} = ( u_r, u_{\varphi}, u_z)$ in
 cylindrical
coordinates ($r$, $\phi $, $z$) is such that
$l=ru_{\phi }$ is constant.  We assume
the polytropic equation of state,
\begin{equation}P={\cal K}\rho ^{\gamma}.\label{1} \end{equation}
Here $P$ is the pressure, $\rho $ is the  density, ${\cal K}$ and $\gamma $ are
the polytropic constant and adiabatic index, respectively.
The sound speed, $c_s$, is then a simple function of the density and
can be written in the form
$$c_s^2={dP \over d\rho }={\cal K}\gamma \rho ^{\gamma -1}.$$

The equation of motion governing the steady state may be written
\begin{equation}{\mbox{\boldmath$\omega$}}\times {\bf u}=
-\nabla{\cal B},\label{2DM}\end{equation}
where ${\mbox{\boldmath$\omega$}}$ is the vorticity and the Bernoulli function
$${\cal B}=
{1\over 2}{\bf u}^2 + {c_s^2 \over \gamma -1} +\Psi,$$
with $\Psi$ being the gravitational potential.
For this we adopt the 'pseudo Newtonian potential'
 (Paczy\'{n}ski and Wiita, 1980)
given by
$$ \Psi = - {GM \over \sqrt{r^2+z^2}-r_G } , $$ where
$G$ is the gravitational constant, $M$ is a mass of the central object and
$r_G$ is the gravitational radius.
 Paczy\'{n}ski and Wiita (1980) emphasized that  all  relatavistic effects
which
are relevant to the accretion process near the inner edge of a  disk can be
reproduced in the Newtonian formalism
assuming the pseudo-Newtonian potential given above.

{}From equation (\ref{2DM}) it follows that ${\cal B}$ is constant on stream
lines and may be specified to vary arbitrarily from stream line
to stream line.
The fact that it is possible to specify an arbitrary function in this
way, being associated with a non zero $\varphi$ component of vorticity,
results in a degree of arbitrariness in possible steady state solutions.
\subsection{The zero vorticity case}
In this paper we shall consider the case when ${\cal B}$  is constant. This
corresponds
to zero vorticity.
This is the simplest assumption that can be made about ${\cal B}$
and it is also the one which is most likely to result in models
that can be approximately described by one dimensional vertically averaged
ones.

It must be noted  that the ability to specify
 the Bernoulli function arbitrarily on stream lines
introduces a high degree of arbitrariness into the two dimensional models.
In principle one might for example be able to  construct
models with flow concentrated towards the disc surface (Paczy\'{n}ski, 1980)
or towards the equatorial plane (Paczy\'{n}ski and Abramowicz, 1982) but
such solutions would have a non zero component of vorticity in the
$\varphi$ direction which can only be specified as an entry condition for the
flow under the assumption of the lack of importance of viscosity we make in the
treatment of the inner region.

When the vorticity is zero,
the meridional component of the velocity field may be expressed as
the gradient of a potential, $\Phi,$ such that
$$( u_r , u_z ) = \left ({\partial \Phi}/{\partial r} , {\partial
\Phi}/{\partial z}\right). $$
For a steady flow, we also have the continuity equation, which for this
flow reads
\begin{equation}\nabla (\rho \nabla \Phi) =0 .\label{4} \end{equation}
This may be combined with  the Bernoulli equation,
$${1\over 2}(\nabla \Phi)^2 + {c_s^2 \over \gamma -1}
 +\Psi + {l^2 \over 2 r^2} = {\cal B}={\rm {constant}}$$
to give a single second order partial differential equation for
the velocity potential $\Phi:$

\begin{equation}\nabla ^2 \Phi - {1 \over c_s^2} \left(\nabla (\Psi_T) + \nabla
\left[ {1 \over 2}(\nabla \Phi)^2\right]\right) \cdot \nabla \Phi =0,
\label{6}\end{equation}
where the total centrifugal plus gravitational
potential is $$\Psi_T=\Psi +{l^2\over 2r^2}.$$
In addition
 the density of the flow can be found from the Bernoulli equation which reads
\begin{equation}\rho =
\left[ {\gamma - 1 \over {\cal K}\gamma }({\cal B} - \Psi_T - {1
\over 2} (\nabla \Phi)^2)\right]^{1\over \gamma -1}. \label{BERN}\end{equation}

In order to discuss possible solutions of equation(\ref{6}),
we first consider properties of the potential $\Psi_T$ for constant specific
angular momentum $l.$

 The equipotential surfaces  have a saddle point or cusp if $l$ is in the range
given by
 $l_{mb}<l<l_{ms}$,
where $l_{mb}$ and $l_{ms}$ are the values of the specific angular momentum at
the marginally bound and marginally stable circular particle orbits
respectively
( Abramowicz et al. 1978).
In order for accretion to take place it is expected that $l$ be in the above
range. In order to construct a steady state solution
we first define a specific angular momentum in the above range and use an
equipotential surface as a reference surface in order to construct
a sonic surface.
\subsection{The sonic surface}
The accretion flow  is expected to be transonic, starting at small subsonic
speed at large radii and accelerating to attain the  sonic speed
along the sonic surface, being the generalization of a critical point in one
dimension.
In general a unique form  for the sonic surface cannot be constructed in
advance of knowing the solution, even if the other bounding surfaces
are known (see Anderson, 1989). However,
construction turns out to be simple in the
special case when  the flow is normal to the surface, (see Anderson, 1989
for reasons why this might be a preferred case).
A surface of this form may be constructed by first selecting a bounding
equipotential surface through which the sonic surface passes orthogonally.
We suppose this to be a zero density surface where the flow approaches
zero velocity, the sound speed being zero.
If the equation of the sonic surface
 with the above properties is $r = r(z),$ on the surface we must have

$${dr\over dz}=-{\partial \Phi\over \partial z}/{\partial \Phi\over \partial
r}.$$ Differentiating this equation with respect to $z$ and using equations
(\ref{6}) and (\ref{BERN}), together with the fact that on the surface
$u_r^2+u_z^2 = c^2_s,$ we find
 the following second order differential equation for the sonic surface:
$${{d^2r \over dz^2}\over  1+\left( {dr \over dz} \right) ^2}=
-{\left({\partial \Psi_T \over \partial r} -{
\partial \Psi_T \over \partial z}
{dr \over dz}\right) \over 2{\gamma -1 \over \gamma +1}({\cal B} -\Psi_T)}
 +{1 \over r}.$$
Here ${\cal B}$ is constant and equal to the value of $\Psi_T$
 on the bounding
 equipotential surface.
The sonic surface that passes orthogonally through the chosen bounding
zero density surface as well as the equatorial plane can be
constructed uniquely by solving this equation.
The zero density surface of the torus, sonic surface, the equator
$(z=0)$ and a bounding surface $r=$ constant on which the inflow velocity
is small are taken to define  the flow domain.
We assume  reflection symmetry in $z$ for the flow below the equator.
At the sonic surface  we set the outflow
velocity equal to the sound speed and  normal to the surface.

In this way the parameters defining a solution are the selected
constant specific angular momentum and the bounding zero density
surface  .
Alternatively one may use the point on the equatorial plane through
which the sonic surface passes (identified with the critical point in one
dimensional flows) $r=r_*,$  and the value of $GM/(r_*c^2_s)$ there  as the
parameters defining a model. The solutions can be scaled to give any mass
accretion rate because they are invariant
 to an arbitrary scaling of the density.

\section{Numerical method for the two dimensional tori}
In order to solve the equilibrium equation for $\Phi $ and at the same
time  find the velocity of the fluid, we
rewrite equation (\ref{6}) in the form of a diffusion type equation:
$${\partial \Phi \over \partial t}=D(r,z,t)\left[ \nabla \left( \rho \nabla
\Phi \right) \right] , $$
where $D(r,z)$ is a diffusion coefficient, which in general can be a function
of $r , z $ and $t$. The form of this can be chosen for numerical convenience
subject to the requirement of numerical stability.
We solved this equation in the elliptic (subsonic)  regime of the computational
grid by
solving the diffusion equation as an initial value problem, integrating forward
until
a steady state was achieved. A two stage procedure was used. During the first
stage,
the equations were advanced with $\rho$ fixed
at the value it would have were the flow  sonic.
Once a steady state had been reached, the
evolution was continued for the second
 stage
during which the density was calculated from equation(\ref{BERN}) until a final
steady state was attained. It is required to specify boundary  conditions on
the
boundary of the computational domain.  These were that the velocity is sonic
and
normal to the sonic surface, that the inflow be normal at the exterior
boundary,
that the flow be tangential at the bounding equipotential surface and symmetry
with respect to reflection in the equatorial plane. The solution is
undertaken on an equally spaced computational grid with between 40 and
50 grid points in the subsonic region of the flow on the equatorial plane.
 It is
straightforward apart from the complication that because the density and
therefore the sound speed go to zero  somewhat interior to the bounding surface
that would occur if there was hydrostatic equilibrium, there must be a
supersonic transition slightly interior to this surface. But note that this
transition occurs at low absolute speeds. The region between this other sonic
surface and the bounding surface is small and we treated it by not allowing the
density to fall below the value taken on when the flow is sonic. In this way a
transition to a hyperbolic region is prevented. However, the location of this
new bounding sonic surface could be found in this way and the solution
continued into the hyperbolic domain by initial value techniques. As this
region
was small, having a negligible effect on the interior,
and in reality it  would depend on details of the proper
boundary condition, we
 did not calculate it in detail.

Before discussing the solutions for two dimensional
 tori that were obtained
 numerically,
we describe the one dimensional models which  correspond to them. In this
regard
we note that the full disk can only be approximated as a constant angular
momentum torus in the innermost region. Further out there must be material
with a higher specific angular momentum and viscous effects to allow it to
accrete. With reference to one dimensional disc models we shall argue below
that
once the two parameters $r_*$ and $GM/(r_*c^2_s)$ are specified and
 if the angular
velocity gradient at the critical point corresponds to constant specific
angular momentum, there is a unique standard " $\alpha$" viscosity parameter
corresponding to the torus.

\section{Basic equations for one dimensional flow}
The basic equations for one dimensional flow with the time dependence retained
are the equation of motion
\begin{equation}{\partial u\over \partial t}+u{\partial u\over \partial r}
=-{1\over \Sigma}{\partial P\over \partial r}+r\Omega^2
-{\partial \Psi\over \partial r}.\label{MOT}\end{equation}
Here $u$ is the radial velocity, $\Sigma$ is the surface density and $P$ is now
the vertically integrated
pressure which we shall assume to be a function of both
$\Sigma$ and $r.$ The latter dependence on $r$
may come about through the process
of vertical averaging (see below).
A local sound speed $c_s$ is then defined through
$$c_s^2=\left({\partial P\over \partial \Sigma}\right)_{r}.$$
The continuity equation is
\begin{equation}{\partial \Sigma \over \partial t}
+u{\partial \Sigma\over \partial r}=
-{\Sigma\over r}{\partial (ur)\over \partial r}.\label{CON}\end{equation}
The specific angular momentum $l(r,t)=r^2\Omega(r,t)$ satisfies
\begin{equation}{\partial l\over \partial t}+u{\partial l\over \partial r}
={1\over (r\Sigma)}{\partial (r^2 \tau_{r\varphi})\over \partial
r}.\label{ANGMOM} \end{equation}
Here $\tau_{r\varphi}$ is the $r \varphi$ component of the
viscous stress tensor.
This is usually taken to be given by
$\tau_{r\varphi}= \tau_{r\varphi 0} =\nu \Sigma r (\partial \Omega /
\partial r)$, with $\nu$ being the kinematic viscosity. For the Shakura and
Sunyaev (1973) $\alpha$ prescription, we have
 $\nu=
\alpha c_s^2/\Omega_K,$ $\Omega_K$ being the circular orbit frequency.
\subsection{The question of causality}
In an important paper, Narayan (1992) has pointed out that
the conventional formulation of viscosity as a diffusion process
allows propagation of information at infinite speed and so could lead
to unphysical results in transonic flows which attain velocities
exceeding the maximum velocity of the information carriers which
produce the viscosity. Unphysical results might arise through
unrealizable diffusive communication between different parts of the fluid.

Using an ilustrative model of the diffusion process with a finite
propagation speed, Narayan (1992) concludes that the $r\varphi$
stress should vanish at a sonic/critical point, where $u^2 = c_s^2,$ in a one
dimensional accretion flow. Here we provide a different but  similar example,
within a Newtonian framework, which uses
all the basic equations. The full set of equations is of hyperbolic type
and so information propagates at  finite speeds. The example shows that
for flows in which the angular velocity and specific angular momentum gradients
are of opposite sign, the viscous stress is, if anything increased, rather than
reduced at the critical point. Furthermore, when the
specific angular momentum gradient is very small when the radial velocities are
significant,
such as might be expected for the flows considered here, the corrections due to
finite propagation effects become small so that the conventional Shakura and
Sunyaev (1973) treatment should be reasonable.
However, constraints arising from causality considerations may be important
when the question of the uniqueness of steady state solutions is discussed.
 When the angular velocity gradient is very large, the conclusions
derived from this simplified model are similar to
those of Narayan (1992).

The basic equations of our simplified model are
equations (\ref{MOT}), (\ref{CON}) and (\ref{ANGMOM}) together with an equation
for $\tau_{r\varphi}$ of the form
 \begin{equation}{\partial \tau_{r\varphi}\over \partial t}
+u{\partial \tau_{r\varphi}\over \partial r}=
{(\tau_{r\varphi 0}-\tau_{r\varphi})\over \tau}.\label{RELX}\end{equation}
This states that the stress relaxes locally to its equilibrium value
on a relaxation timescale $\tau$ which may be an arbitrary function of the
state variables but not their gradients.
We stress that this formalism has been introduced to ensure that information
propagates at a finite speed in the system rather than to provide a
realistic formulation of viscosity. It  is thus adequate to demonstrate
that zero viscous stress at the sonic point cannot be inferred
from any limitation in the propagation speed of information.
Equations (\ref{MOT}), (\ref{CON}), (\ref{ANGMOM}) and (\ref{RELX})
form a system of four simultaneous first order partial differential equations
which may be written in the form:
\begin{equation}
{\partial U_i\over\partial t}
+\sum^{j=4}_{j=1}A_{ij}{\partial U_j\over\partial r}+S_i=0.\
(i=1,2,3,4)\end{equation}
Here $U^T =(\Sigma ,u ,l, \tau_{r\varphi}),\ S_1=\Sigma u/r,\
S_2={1\over \Sigma}\left({\partial P\over \partial r}\right)_{\Sigma}-r\Omega^2
+{\partial \Psi\over \partial r},\ S_3=-2\tau_{r\varphi}/\Sigma,\
S_4=(\tau_{r\varphi}+2\nu\Sigma lr^{-2})\tau^{-1}.$ The diagonal elements of
the matrix $(A)$ are all equal to $u,$ $A_{12}=\Sigma,\  A_{21}=c_s^2/\Sigma,\
A_{34}=-r/\Sigma,\  A_{43}=-\nu\Sigma/(\tau r),\ $ with all other elements
being
zero.
The system is hyperbolic if, as for our set of equations, the eigenvalues of
$A$
are all real. The characteristics are rays with propagation speeds given by
these eigenvalues. In our case these are the two sonic speeds $u+c_s, u-c_s$
and the two viscous speeds $u+c_v,\ u- c_v,$ with
$c_v=\sqrt{\nu/\tau},$ being the speed of propagation of viscous information.
Therefore information cannot propagate faster than a finite speed in the model
system. But note that the viscous speed, which we assume to be comparable to
the sound speed, becomes infinite when the relaxation time is zero.
When $c_v = c_s,$ we have $\alpha = \Omega_K\tau.$
\subsection{The steady state}
We now examine the steady state in which ${\partial \over\partial t}\equiv 0.$
 The continuity equation gives on integration
\begin{equation}
2\pi\Sigma ru = -c_1,\label{CONIN}
\end{equation}
$c_1$ being the constant accretion rate.
Equation (\ref{ANGMOM}) gives on integration
 \begin{equation}
c_1l(r)+2\pi r^2 \tau_{r\varphi}= c_1c_2,\label{AM}
\end{equation}
where $c_2$ is the constant  rate of flow of angular momentum through a circle
of radius $r.$
Equation (\ref{RELX}) gives
\begin{equation}\tau_{r\varphi} =
       \nu \Sigma r (d\Omega / dr)
-\tau u{d \tau_{r\varphi}\over d r}.\end{equation}
{}From the above two equations it follows, after differentiating the first and
using the result to eliminate  ${d \tau_{r\varphi}\over d r},$ that:
\begin{equation}\tau_{r\varphi} =
      {\nu \Sigma r \left((d\Omega / dr)-(u^2 / (c^2_v r^2))(dl/dr)\right)
      \over(1-(2u\tau/ r))}.\label{CRAP}\end{equation}
This is the Shakura and Sunyaev (1973) viscosity prescription modified by
additional terms involving the radial velocity.
We first remark that for $u\sim -c_s,$ and $\tau\le\Omega^{-1},$
$|u\tau/r|\le  H/r,\ H$ being the disk thickness. Thus for most purposes
the denominator in equation (\ref{CRAP}) may be approximated as unity
 (it could also be absorbed by a redefinition of $\alpha$ incorporating
radial dependence).

We note that corrections arising from propagation effects are small
if $|dl/dr|$ is small such the flow has almost constant specufic angular
momentum even when $u^2=c_s^2.$
Furthermore the viscous stress is increased if the gradients of angular
velocity and angular momentum are of opposite sign.
 But note that an interesting
situation arises
 at a sonic point in the physically reasonable case when $c_v =c_s.$
Equation (\ref{CRAP}) may be written
\begin{equation}\tau_{r\varphi} =
      {\nu \Sigma r \left((1-(u^2 / (c^2_s ))(d\Omega/dr)-(2\Omega u^2 )/
(rc^2_s)\right)
      \over(1-(2u\tau/ r))}.\label{CRCP}\end{equation}
{}From this it follows that at a sonic point:
\begin{equation}\tau_{r\varphi} = -
      {2\nu \Sigma\Omega
      \over(1+{2c_s\tau\over r})}.\label{CRBP}\end{equation}
We see that the stress is independent of the angular velocity gradient
and takes on the same form as it would do in the case of constant specific
angular momentum. Equation(\ref{CRBP}) shows that the angular momentum flux is
then fixed at the sonic point by the magnitude of $\Omega.$

This is in contrast to the situation when the viscosity prescription
with $u=0$ is used. Then the angular momentum flux may be adjusted by
altering the angular velocity gradient at the sonic point, albeit
by a small amount if $\alpha$ is small.
Within this prescription, when $d\Omega/dr=-2\Omega/r,$ the angular momentum
flux
is also fixed by $\Omega$ at the sonic point and the standard $\alpha$
prescription is essentially unmodified by finite propagation effects.
We shall refer to
this condition at the sonic point as the natural boundary condition.

In fact equation (\ref{CRCP}) implies the existence of a critical point at the
sonic point and that equation (\ref{CRBP})
should be satisfied there to avoid a singularity and that indeed freedom to
specify the angular velocity gradient should be lost. However,
the technicalities of  additional  critical point
analysis ( scarcely justified by the
sophistication of the model ) can be avoided
by noting that for our solutions, when the natural boundary condition
is used, equation (\ref{CRBP}) is essentially
satisfied in any case and the angular
momentum profiles are so flat that limitations due to finite propagation
effects may be ignored. Note too that this is reasonable on physical
grounds because viscous effects are small in any case when the angular
momentum profile is flat.

 For the simple model discussed here, we have seen that
when the viscous propagation speed is equal to the sound speed, freedom to
adjust the angular momentum flux by varying the angular velocity gradient at
the
critical point is lost. This flux is always the same as when the natural
boundary
condition is used. We shall see that this is important for the issue
of uniqueness of steady state solutions.

When the mass flow rate, $\alpha$
and the enropy of an adiabatic steady state solution are specified and the
natural boundary condition is used, the solution is apparantly
unique. However, if there is freedom to adjust the angular momentum flux by
varying the angular velocity gradient at the critical point, as
might arise if $c_v > c_s,$ the solution is no longer unique (see below).

Finally, we comment that when the angular velocity gradient is {\it very}
steep,
we have approximately
\begin{equation}\tau_{r\varphi} =
      {\nu \Sigma r \left(1-{u^2\over c_s^2}\right)(d\Omega / dr)
      \over(1-{2u\tau\over r})}.\label{CRDP}\end{equation}
This is of a similar form to that given by Narayan (1992).

In the numerical work presented below, we adopt the standard Shakura and
Sunyaev (1973) form: $\tau_{r\varphi} =\tau_{r\varphi 0},$ noting that
we expect modifications due to finite propagation effects to be
negligible, when the natural boundary condition is used.

For the steady state, the equation of motion reads:
\begin{equation}
u{du \over dr} = -{1 \over \Sigma}
{dP \over dr}- {d\Psi \over dr }+{l^2 \over r^3},
\end{equation}
The vertically integrated pressure may be taken to obey a polytropic equation
of state  $P=K\Sigma
^{1+1/ n},$ $n$ being the polytropic index.
The polytropic factor  $K$ may be taken to be constant, or a function of $r$
to reflect vertical averaging.
This dependence can be found from the scalings
for an adiabatic equation of state:
$P\propto H\rho^{\gamma},\ \Sigma\propto \rho H,\ $ and
$\Omega_K^2H^2\propto c_s^2\propto \rho^{\gamma-1},$ with
$\Omega_K=\sqrt{GM /( r(r-r_G)^2)}$
being the Keplerian angular velocity for circular orbits.
Eliminating $H$ and $\rho$ from these gives
$$K=K_0\Omega_K^{{2(\gamma-1)\over(\gamma+1)}},$$ $K_0$ being constant
and $1+1/n=(3\gamma-1)/(\gamma+1).$
Combining the equation of motion with the continuity equation and the
above equation of state gives
\begin{equation}
\left( u - {c_s^2 \over u }\right) {du \over dr} =
-{GM \over (r-r_G)^2} + {c_s^2 \over r}\left(1-{nr\over (n+1)K}{dK\over
dr}\right) + \Omega ^2 r.\label{EM} \end{equation}
This exhibits a critical point when the flow speed reaches the sound speed.
At such a point the right hand side of equation (\ref{EM}) must vanish.
The nature of the critical point when $K$ is constant has been discussed
by Papaloizou and Szuszkiewicz (1993) who find that for physically acceptable
solutions it is of saddle type.
\subsection{Dimensionless form of the equations}
In order to discuss solutions of the steady state equations further,
we adopt the usual "$\alpha$" viscosity:
$$\tau_{r\varphi}=\tau_{r\varphi 0}=\nu \Sigma r (d\Omega / dr),$$
 with $\nu= \alpha c_s^2/\Omega_K.$
We use this together with equations (\ref{CONIN}), (\ref{AM}) and
(\ref{EM}).

We find  it convenient to  adopt as units of radius and velocity $r_*$ and
$u_*$ being the radius at the critical point and the absolute magnitude of the
velocity velocity there, respectively. Then we define the dimensionless
variables
$v=-u/u_*$, $x=r/r_*$, $\omega = \Omega r_*/u_*,$ $\omega_K = \Omega_K r_*/u_*$
 and $x_G=r_G/r_*.$
Note that because we are dealing with inflow $v$ is positive.
The square of the dimensionless
sound speed is  $\tilde{c_s^2}=\tilde{K}/(vx)^{1/n},$
with $\tilde{K}= K(x)/K(1).$
In addition we define  $\tilde{c_2}=c_2/(u_*r_*)$,
and $m=GM/(r_*u^2_*)$ as the Bondi parameter.

It is possible to derive a single nonlinear second order differential
equation for $\omega$ from the above equations which may be written

\begin{equation}
{\cal F}={1\over n_1}\left( v^{1-1/n} - {\tilde {K} \over x^{1/n}v^{1+2/n}
}\right) {dv^{n_1} \over dx} + {m\over (x-x_G)^2} - {\tilde{K}\over
v^{1/n}x^{n_1}} \left(1-{x\over n_1\tilde{K}}{d\tilde{K}\over dx}\right) -
\omega^2
x=0,\label{DEM}\end{equation} where, $n_1 = 1+1/n\ $ and
\begin{equation}
 v^{n_1} = {\alpha\tilde{K}  \over \omega_K x^{1/n- 2}
(\tilde{c_2}-\omega x^2)}{d\omega \over dx}.\label{VI}\end{equation}
At the critical point where $x = v = \tilde{K} =1,$ equation (\ref{DEM})
requires that the dimensionless angular velocity $\omega$ be given by
  $$\omega_* =\sqrt{{m \over (1-x_G)^2} -
1+{n\over (n+1)}\left({d\tilde{K}\over dx}\right)_{x=1} }.$$
Equation (\ref{VI}) then gives
\begin{equation}\tilde{c_2} = \left({\alpha \over \omega_{K}}{d\omega \over dx
}\right)_{x=1} +\omega_*. \label{JF}\end{equation}
We have also found it convenient to work in terms of a quantity $c_N$
related to $\tilde{c_2}$ through

\begin{equation}\tilde{c_2}
 = -{2 c_N \alpha \omega_* \over \omega_{K}(1)}
+\omega_*.\label{JBJ}\end{equation}
Thus when the natural boundary condition is used $\tilde{c_N} = 1.$

\subsection{Parameters specifying a solution}
Equation (\ref{DEM}) after use of (\ref{VI}) constitutes
 a second order differential
equation for $\omega.$ After specification of the Bondi parameter $m, x_G,$
$\alpha$ and $\tilde{c_2},$ both $\omega$ and its first derivative are
specified at the critical point. For the critical points considered here
these conditions specify a unique solution that is subsonic for $x>1$
( Papaloizou and Szuszkiewicz, 1993). In order to be physically acceptable
this solution must match onto a Keplerian or near Keplerian disk at some
specified large radius. In general this requires that one of the parameters
such as $\alpha$ be adjusted as an eigenvalue in order to fulfill this
requirement. In this way we see that specification of $m, x_G$ and
$\tilde{c_2}$
determines a value of $\alpha$
required to satisfy the boundary conditions.

However the parameter $x_G$ which can be regarded as specifying the location
of the critical point may also be varied, other parameters being kept
 fixed,  so determining this quantity as a
function of $\alpha.$

On physical grounds we expect that for adiabatic solutions of the type we are
considering, we should be able to specify the accretion rate and the entropy
with some degree of arbitrariness. The solutions can always be adjusted to
give any accretion rate by scaling the constant $c_1,$ while the Bondi
parameter $m$  may be used to adjust the entropy.
If these are determined in this way, the above discussion shows that when
$\tilde{c_2}$ is fixed the location of the critical point is then determined
as a function of $\alpha.$

On dynamical grounds we expect the solutions considered here to be of nearly
constant specific angular momentum when there are significant velocities and
the
discussion of the requirement of a finite propagation speed of viscous
information given above then suggests that $\tilde{c_2}$ be chosen to satisfy
the natural boundary condition such that $$\left({d\omega \over dx
}\right)_{x=1} =-2\omega_*.$$
Under these conditions $\tilde{c_2}$ is no longer free and we expect that
for a given accretion rate, entropy  and $\alpha,$ the solution
together with the location of the critical point are specified uniquely
{\it at least locally in parameter space}. However, if we allow
freedom in the specification of $\tilde{c_2}$ this is no longer precisely so.
For a given $m$ and $x_G$ there is a possible range of $\alpha$ corresponding
to the range in $\tilde{c_2}$ or equivalently for a given $\alpha$
there is a possible variation in $x_G$ or the location of the critical point.

Variation of $\tilde{c_2}$ causes the angular velocity gradient to differ from
that appropriate to constant specific angular momentum at the critical point.
Because the dynamics requires that the angular momentum profile does not differ
much from a constant, this situation is restored in a fairly narrow boundary
layer when the boundary condition for $\tilde{c_2}$ differs from the natural
one. In practice, for a fixed $m,$ the variation
in $\alpha$ allowed for a given $x_G,$ or
the variation in $x_G$ found for a given $\alpha$ consequent on
varying $\tilde{c_2}$ by a maximum reasonable amount, is found to be small.

Adopting the natural boundary condition, the above discussion indicates
that specification of $m$ and $\alpha$ results in a unique critical
point location given through $x_G.$

Such a solution can then be matched
to a two dimensional solution with constant specific angular momentum
which has the same value of $m$ and which has a sonic surface which
passes through the equatorial plane at the same location as the critical point
in the one dimensional case.

\subsection{Numerical methods for the one dimensional case}
Numerical solutions in the one dimensional case have been generated
by two methods. In the first, the domain between the critical point
and the outer boundary  was divided into $N$ equally spaced zones, with
values of $N$ up to 18000 having been used.
Equations
(\ref{DEM}) and (\ref{VI}) were then written in a centered finite difference
form. On specification of $c_N , m,  x_G $ and $\alpha,$ the solution can be
found by stepping outwards from the critical point. However, as discussed
above,
the condition  that the angular velocity match on to the Keplerian value at
some
outer boundary radius, in general cannot be satisfied unless $\alpha$
takes on a specific characteristic value.
This method was most satisfactory when the critical point was inside
$3r_G$ and $m$ was not too large. When $m$ is large, the disk thickness
is very much less than the radius, and the
equations are stiff. This has the effect that with the above method
it is awkward to extend the solution to very large outer boundary radii.
However, the characteristic value of $\alpha$ corresponding to
given $x_G$ and $c_N$ is very accurately determined.

An alternative approach is to use a relaxation method (Papaloizou and
Szuszkiewicz 1993) such as for example solving the diffusion type equation
derived from equation(\ref{DEM})

\begin{equation}{\partial \omega \over \partial t}={\cal D}{\cal
F}\end{equation}
where ${\cal D}$ is a diffusion coefficient which can be a function
of $x$ and $t$ and which can be chosen for numerical convenience.

This can be solved in a computational domain extending from the critical point
to the outer boundary. In this case $\omega$ is specified at the domain
end points. For fixed $\alpha, m $ and $x_G$ this is enough to specify the
solution but then $\tilde{c_2}$ is determined in the course of
solution through equation
(\ref{VI}). This has the disadvantage that
 in order for a solution to exist, once
$x_G$ and $m$  have been specified, $\alpha$
 must be in the range given through the
maximum permitted variation of $ \tilde{c_2}$ (see below).
This method which does not have the
  stiffness difficulties of the first
 was found
to work best when the critical point was outside $3r_G.$ Some solutions
obtained by the first method were checked using this method.

\section{Numerical Results}
We have constructed two dimensional model tori
with  different values of the  constant specific  angular momentum and
 vertical thickness.
Each torus  is characterized by a value of $m$ and
$x_G$ given in Table 1.
 For each of these models
the outer boundary was taken at $r = 4r_G$ which was far enough out
for the inflow velocity to be small there, making it seem reasonable
that the disk can be regarded as essentially static beyond.

The results we describe below are qualititatively similar for all of the models
we examined so we shall focus on two examples for illustrative purposes.
We show the velocity and density structure appropriate to
models 3 and 12 in figures 1 and 2 respectively.
 Model 3 represents a  moderately thick
disc. The inner edge of this disc, where the radial inflow reaches
the sound speed, is located at the radius $\approx 2.5r_G$.
The maximum density occurs around $3.5 r_G$.
Model 12 is an example of a  genuinely thick accretion disc.
Its inner edge is close to $2.05r_G$ and the density maximum is outside
$4r_G$.

For each of the two dimensional models we have constructed
a one dimensional viscous equivalent with the same values of $m$
and $x_G.$ As argued above if $c_N,$ which controls the
angular velocity gradient at the critical point is fixed, there is
a unique value of $\alpha$ for which the angular velocity
takes on the Keplerian value at some outer radius, here
taken to be at $10 r_G.$
We denote the value of $\alpha$ corresponding to $c_N=1,$
 the natural boundary condition,  by $\alpha_{max}.$
 In addition we
denote the value of $\alpha$ corresponding
to $c_N =0 $ or zero angular velocity gradient at the critical
point by $\alpha_{min}.$

\begin{table*}
\caption{The various two dimensional model calculations: $m$ is the Bondi
parameter measured where the sonic surface intersects the equatorial plane at
$r=r_*.$ The equipotential surfaces have a cusp at $r = r_G r_{eq}$,
$h$ is the ratio of the vertical distance between the equatorial plane
and the zero density surface
of the torus, to the radius
and $x_G =
r_G/r_*.$
 In addition we show $\alpha_{max}$ and $\alpha_{min},$
the values of $\alpha$ appropriate to
 one dimensional models, with vertical
 averaging taken into account in the equation
 of state,  with the same
$x_G$ and $m.$ The former value corresponds to the natural boundary condition
while the latter corresponds to zero angular velocity gradient at the critical
point. For each one dimensional model, the exterior boundary was taken at $10
r_G.$} \vspace{0.5cm}

\begin{tabular}{cccccccc}

Model & $m$   & $x_G$   & $r_{eq}$   &  $h$    & $\alpha_{max}$    &
$\alpha_{mi
n}$\\
1 & 302& 0.34 & 2.9 &0.1 &$6.03\times 10^{-2}$& $2.98\times 10^{-2}$ \\
2 & 83 & 0.39 & 2.9 &0.2 &$1.26\times 10^{-2}$& $1.07\times 10^{-2}$ \\
3 & 75 & 0.39 & 2.7 &0.2 &$1.40\times 10^{-2}$& $1.17\times 10^{-2}$ \\
4 & 21 & 0.42 & 2.7 &0.4 &$1.81\times 10^{-2}$& $1.50\times 10^{-2}$ \\
5 & 63 & 0.43 & 2.4 &0.2 &$3.65\times 10^{-3}$& $3.48\times 10^{-3}$ \\
6 & 16 & 0.43 & 2.7 &0.5 &$1.58\times 10^{-3}$& $1.34\times 10^{-2}$ \\
7 & 18 & 0.45 & 2.4 &0.4 &$9.01\times 10^{-3}$& $8.18\times 10^{-3}$ \\
8 & 13 & 0.45 & 2.4 &0.5 &$9.94\times 10^{-3}$& $8.97\times 10^{-3}$ \\
9 & 55 & 0.46 & 2.2 &0.2 &$5.67\times 10^{-4}$& $5.63\times 10^{-4}$ \\
10& 51 & 0.48 & 2.1 &0.2 &$2.59\times 10^{-4}$& $2.58\times 10^{-4}$ \\
11& 15 & 0.48 & 2.2 &0.4 &$3.50\times 10^{-3}$& $3.38\times 10^{-3}$ \\
12& 11 & 0.49 & 2.2 &0.5 &$4.02\times 10^{-3}$& $3.86\times 10^{-3}$ \\
\end{tabular} \vspace{1cm}
\end{table*}
The range between $\alpha_{max}$ and $\alpha_{min}$ is spanned
continuously as $c_N$ is varied from $1$ to $0.$ Values of $c_N$ less than $0$
correspond to the situation where the angular velocity gradient
turns over. If variation in $c_N$ is permitted this results
in a range of possible $x_G$ for fixed $m$ and $\alpha,$
(see the above discussion of uniqueness).
We remark
that $\alpha_{max}$ and $\alpha_{min}$ do not normally differ by more
than twenty percent. In addition, other parameters being fixed,
 these values are somewhat
sensitive to the location of the outer boundary and tend to decrease
as this moves outwards. The sensitivity is greater for smaller
values of $m.$ The angular momentum profile interior to $4r_G$ is
very flat with $l\propto r^{\sim 0.04}.$ This is fully consistent
with the idea that viscous effects are weak in this region
and justifies the approximation of constant specific angular
momentum.
The one dimensional models do not depend very much on whether
vertical averaging is taken into account in the equation of state
  ($K=K(x)$) or a straight forward polytropic
equation of state ($K=\rm{constant}$) is used.

Values of $\alpha_{max}$ and $\alpha_{min}$ for some values of $x_G$ and $m$
 are given for the former case in table $2$ and the latter case
in table $3.$

 \begin{table*}
\caption{Values of $\alpha_{max}$ and $\alpha_{min},$
for a sample of one dimensional models with various
$x_G$ and $m$ for the case when vertical averaging was taken into account
in the equation of state. In each case $\gamma =4/3$ and the external radius
was
 taken at $10r_G.$}
\vspace{0.5cm}

\begin{tabular}{cccccccc}

 $m$   & $x_G$   & $\alpha_{max}$      & $\alpha_{min}$\\
10  & 0.32 & $1.67\times 10^{-1}$& $6.24 \times 10^{-2}$ \\
100 & 0.35 & $6.51\times 10^{-2}$& $3.39 \times 10^{-2}$ \\
10  & 0.35 & $1.00\times 10^{-1}$& $5.08 \times 10^{-2}$ \\
100 & 0.40 & $7.45\times 10^{-3}$& $6.71 \times 10^{-3}$ \\
10  & 0.40 & $4.29\times 10^{-2}$& $3.00 \times 10^{-2}$ \\
100 & 0.45 & $2.71\times 10^{-4}$& $2.71 \times 10^{-4}$ \\
10  & 0.45 & $1.44\times 10^{-2}$& $1.25 \times 10^{-2}$ \\
100 & 0.50 & $7.61\times 10^{-6}$& $7.61 \times 10^{-6}$ \\
10  & 0.50 & $3.18\times 10^{-3}$& $3.08 \times 10^{-3}$ \\
\end{tabular} \vspace{1cm}
\end{table*}

\begin{table*}

\caption{Values of $\alpha_{max}$ and $\alpha_{min},$
for a sample of one dimensional models with various
$x_G$ and $m$ for the case when vertical averaging was not taken into account
in the equation of state.
In each case $\gamma = 4/3$ and the external radius was taken at $10r_G.$}
\vspace{0.5cm}

\begin{tabular}{cccccccc}

 $m$   & $x_G$   & $\alpha_{max}$       & $\alpha_{min}$\\
10  & 0.32 & $1.23\times 10^{-1}$& $5.51 \times 10^{-2}$ \\
100 & 0.35 & $5.96\times 10^{-2}$& $3.26 \times 10^{-2}$ \\
10  & 0.35 & $7.88\times 10^{-2}$& $4.46 \times 10^{-2}$ \\
100 & 0.40 & $8.01\times 10^{-3}$& $7.23 \times 10^{-3}$ \\
10  & 0.40 & $3.55\times 10^{-2}$& $2.64 \times 10^{-2}$ \\
100 & 0.45 & $4.71\times 10^{-4}$& $4.70 \times 10^{-4}$ \\
10  & 0.45 & $1.31\times 10^{-2}$& $1.17 \times 10^{-2}$ \\
100 & 0.50 & $2.41\times 10^{-5}$& $2.33 \times 10^{-5}$ \\
10  & 0.50 & $3.54\times 10^{-3}$& $3.43 \times 10^{-3}$ \\
\end{tabular} \vspace{1cm}
\end{table*}
It is seen that a range of $7.6 \times 10^{-6} \le \alpha \le 1.7\times
10^{-1}$
 can be spanned by varying $m$ and $x_G.$ The specific angular momentum
  for  one dimensional models ( with vertical averaging taken into account
in the equation of state )
and two dimensional models are compared in figures 3 and 4.
In figure 3 we show the distributions of specific angular momentum
in the model 3  torus and in the corresponding
 one dimensional vertically integrated models
with $c_N$ equal to  zero and one respectively. For $c_N=0$ a  very narrow,
but clearly pronunced boundary layer exists. In this layer, the specific
angular momentum rises steeply, eventually attaining an almost constant value.
In figure 4 we plot the same curves but for model 12.

The radial velocities in the one dimensional models are not
very sensitive to whether $c_N = 0$ or $c_N =1$ is used.
This is illustrated in figures 5 and 6 in which the radial velocities
are plotted for the one dimensional models corresponding to models
3 and 12 respectively for a small range of radii near to the critical point.

In figures 7 and 8 radial velocity profiles
in one and two dimensional cases are compared for models 3 and 12 respectively.
 For the two dimensional tori, we plot
 the radial velocity measured  on the equatorial plane. We plot the radial
velocities for the corresponding one dimensional models with and without
vertical averaging being taken into account in the equation of state.
There is good agreement between all three of the radial velocity profiles
for each model. This indicates
 that the results are not very sensitive
 to the precise mode of vertical averaging.

\section{Discussion}

In  this paper we constructed steady
 state non static accretion tori with
constant specific angular momentum  and entropy. In general there is
some degree of arbitrariness in this procedure because the Bernoulli
integral may be chosen to vary in an arbitrary way on stream lines.
This corresponds to differring specifications for the $\varphi$
component of vorticity which needs to be specified as an entry condition
on the flow. For simplicity we have restricted attention to the case of zero
vorticity.
 In this case we derived an
equation for the velocity potential and solved it to give solutions
appropriate to the  inner transonic region of an accretion disc.

 For the purpose of comparison of these models with models obtained in a  one
dimensional approximation, we considered such models, including
viscosity treated according to the usual Shakura and Sunyaev (1973)
 $\alpha$ prescription.

An important issue is whether specification of the mass accretion rate,
entropy ( or sound speed at the critical point) and the constraint that the
material have the specific angular momentum appropriate to a Keplerian
disc  at some specified outer boundary radius is enough to fix
the location of the critical point for a given value of $\alpha$.

  We   found that if the usual $\alpha$  viscosity prescription
is used in which the viscous stress
is proportional to the angular velocity gradient,
the location of the critical point is not determined uniquely,
although the possible spread in locations might be small.
Note too that previous work by Abramowicz {\it et al} (1988)
used a viscosity prescription in which the viscous stress was simply
proportional to the pressure. In this case one expects a unique
determination for the location of the critical point (at least locally
in parameter space.) The extra freedom in our case arises from
the possibility of varying the viscous stress at the critical point
through changing the angular velocity gradient.

In order to investigate this situation further,
we discussed, using a simple model, the possible
effects of the limitation that viscous information should not be
transmitted at a speed exceeding the sound velocity (Narayan, 1992).
We found that
contrary to some previous work, the viscous stress does not vanish
at the critical point but that it tends to be increased if as in our
models, the angular momentum and angular velocity gradients have opposite sign.
In addition models such as  the ones we constructed with very flat
 specific angular momentum profiles throughout ( ie with no boundary layer)
 are barely
 affected by this propagation constraint.

However, if the speed associated with the propagation of viscous information
 coincides with the sound speed, the ability to change the viscous stress
through varying the angular velocity  gradient at the critical point
together with the lack of uniqueness in determining its location
 may be removed ( at least in our simple model).

In  general we found good agreement between flow
parameters determined from the one and two dimensional models.
When the intersection
point of the sonic surface with the equatorial plane in a two
dimensional model coincides with the critical point in a one
dimensional model and the sound velocities match there
good agreement is found for the radial velocity profile  measured
 on the equatorial plane in the two dimensional case
and the radial velocity profile found in the one dimensional case.

\newpage
{\bf Figure captions}

Figure 1: The density structure (gray scale levels) and the velocity
field (arrows) for the model 3 torus
 with $m=75$ and $x_G=0.39$.
The dashed curve is the sonic surface.

Figuere 2: As in figure 1 but for model 12 which has
 with $m=11$ and $x_G=0.49$.

Figure 3: The specific angular momentum distribution,
given in units of $\sqrt{GMr_G}$, for the model 3  torus (dashed
line) and  corresponding one dimensional
models (dot-dashed and solid lines). The dot-dashed line represents the
solution with the natural boundary condition $c_N=1$ and the solid line
the solution with $c_N=0$.

Figure 4: As in figure 3 but for model 12.

Figure 5: The magnified region near the critical point showing the
comparison between the radial velocities in the one dimensional models
 obtained with $c_N=1$ (dot-dashed
curve) and $c_N=0$ (solid curve) for model 3.

Figure 6: As in figure 5 but for model 12.

Figure 7: The radial velocity  in   units of  the sound speed at
the critical point is  plotted for the model 3 torus
(dashed curve), for the  one dimensional model
with vertical averaging taken into account in the equation of state
 (solid curve) and for the two dimensional polytrope
(dot-dashed curve):

Figure 8: As in figure 7 but for model 12.

\newpage

{\bf References}

Abramowicz, M. A., Czerny, B., Lasota, J-P., \&, Szuszkiewicz, E., 1988,
{\it Ap. J.} {\bf 332}, 646

Abramowicz, M. A., Jaroszy\'{n}ski, M., \& Sikora, M., 1978, {\it
Astron. Astrophys.} {\bf 63}, 221

Abramowicz, M. A., \& Zurek, W., 1981, {\it Ap. J.} {\bf 246}, 314

Anderson, M.,1989,
{\it Mon. Not. R. astr. Soc.} {\bf  239}, 1-17

Chen, X., \& Taam, R. E., 1993, {\it Astrophys. J.}, {\bf 412}, 254

Kato, S. Honma, F., \& Matsumota, R., 1988, {\it Publ. Astr. Soc. Japan},
{\bf 40}, 709

Narayan, R., 1992, {\it Astrophys. J.}, {\bf 394} 261

Paczy\'{n}ski, B., 1980, {\it Acta Astron.}, {\bf 30}, 347

Paczy\'{n}ski, B. \& Abramowicz, M. A., 1982, {\it Astrophys. J.}, {\bf 253},
897

Paczy\'{n}ski, B. \& Wiita, P. J., 1980, {Astron. and Astrophys.}, {\bf 88},23

Papaloizou, J. C. B., \& Szuszkiewicz, E., 1993, in {\it 33rd Herstmonceux
Conference "The Nature of Compact Objects in AGN"} in press

R\'{o}\.{z}yczka, M. \& Muchotrzeb, B., 1982, {\it Acta Astron.}, {\bf 32}, 285

Shakura, N. I., \& Sunyaev, R. A., 1973, {\it Astron. and Astrophys.},
{\bf 24}, 337

\end{document}